\documentstyle[12pt]{article}
\begin{document}

\begin{center}
{\Large\bf Gauge Independent Lagrangian Reduction of Constrained Systems}
\vskip 0.5in
R. Banerjee\footnote{On leave of absence from S.N.Bose National Centre
for Basic Sciences, Calcutta, India. E-mail: rabin@if.ufrj.br}\\[.5cm]
Instituto de Fisica\\
Universidade Federal do Rio de Janeiro, C.P. 68528\\
21945-970 Rio de Janeiro (RJ)\\
Brasil
\end{center}

\begin{abstract}
A gauge independent method of obtaining the reduced space of constrained
dynamical systems is discussed in a purely lagrangian formalism. Implications
of gauge fixing are also considered.
\end{abstract}
\newpage

An important aspect in the hamiltonian formulation of gauge theories
is to obtain the reduced (physical) space comprising the true canonical
variables. This is usually done by fixing a gauge that removes the
unphysical degrees of freedom \cite{1,2}. To avoid the ambiguities and
arbitrariness inherent in the gauge fixing procedure it becomes
desirable to abstract the reduced space in a gauge independent
manner \cite{3}. This also helps in defining a class of admissible or allowed
gauges as those which yield a reduced space that is equivalent,
modulo canonical transformations, to the one obtained in the
gaugeless scheme \cite{4}. Now for a system of constraints in strong
involution (as happens, for example, in abelian gauge theories)
there is a definite gauge independent hamiltonian method of reducing
the degrees of freedom that is based upon the Levi-Civita
transformation \cite{5}. This idea has been exploited to obtain the reduced
phase space  of several models \cite{4,6,7}.

In this paper, using certain results from the theory of differential
equations, a purely lagrangian approach for obtaining the reduced
space in a gauge independent manner will be discussed.
Consequently it provides a lagrangian realisation of the Levi-Civita
reduction process \cite{4,5,6,7}.  Moreover the
proposed method is more direct and does not require the Dirac \cite{1}
algorithm for computing the constraints or their classification into
first and second class, which is an essential perquisite for the 
hamiltonian Levi-Civita method. In this sense the present analysis is similar
in spirit to the symplectic approach \cite{8} based on Darboux theorem but,
contrary to it, does not need a first order lagrangian as the
starting point. Indeed both first and second order systems will be
discussed here on an identical footing. The effect of gauge fixing is
also considered. It is shown that the hamiltonian formalism admits a
wider class of allowed gauges compared to the present lagrangian
formalism. This exercise illuminates, if not settles, the debate \cite{4,9,10}
regarding the simultaneous imposition of the axial $(A_3 \approx 0)$
and temporal $(A_0 \approx 0)$ gauges in pure electrodynamics.

         From the theory of differential equations unsolvable with
respect to the highest derivatives, it is possible to express the
lagrange equations for second order systems with variables $v$ by an
equivalent set of independent equations \cite{3},
\begin{eqnarray}
\ddot p & =& \Theta (p, \dot p, q, \beta, \dot {\beta}, \ddot
\beta) \nonumber \\
\dot q & =& \Phi (p, \dot p, q, \beta, \dot {\beta}) \nonumber \\
r & =& \Psi (p,  q, \beta) \label{1}
\end {eqnarray}
where $v = (p, q, r, \beta )$ and $\Theta, \Phi, \Psi $ are some
functions of the indicated arguments. In a nonsingular theory $q, r,
\beta $ are absent so that there is an unconstrained dynamics with  
$ \ddot p = \Theta (p, \dot p)$. For singular theories
the last two equations
of (\ref{1}) represent the constraints. Now recall that the lagrange equations
were derived by a variational principle on the assumption that all
$v, \dot v$ were free. Since the constraints impose certain
restrictions on $v, \dot v$, it is essential that these keep the set
of equations (\ref{1}) unmodified, or internal consistency is lost.
Consequently time derivatives of the constraints must vanish by
virtue of these equations. This implies that the complete
constraint sector is contained in (\ref{1}). It avoids the Dirac (\cite{1})
algorithm of
iteratively generating this sector in the hamiltonian
formalism. Note also that the absense of any equation for $\beta$ indicates a
possible degeneracy in (\ref{1}). 

    The idea is now to pass from the constrained $v = (p, q, r,
\beta)$ to the unconstrained $v = p$ by removing $q, r, \beta$. The
variable $r$ can be trivially eliminated in favour of $p, q, \beta$. 
In the physically interesting gauge systems the
constraints are implemented by a lagrange multiplier whose time derivative,
therefore, does not appear in the lagrangian. This multiplier is
identified with $q$ which can thus be removed in favour of $p, \beta
$ by using (\ref{1}). The lagrangian in the reduced sector is now a
function of $(v, \dot v; v= p, \beta) $. By evaluating the lagrange
equations in this sector it is possible to recognise $\beta $ as the
variable that does not occur in these equations. 
 This suggests a specific polarisation of the reduced variables
that isolates $\beta$ eliminating it
automatically from the lagrangian and its
final unconstrained form is obtained. The physical hamiltonian is now
found by performing the standard Legendre transformation.

   The same analysis is now applied for first order systems. The form
of the lagrange equations analogous to (\ref {1}) is given by,
\begin{eqnarray}
\dot p & =& \Phi (p,  \beta, \dot {\beta}) \nonumber \\
q & = & \Psi (p, \beta ) \label {2}
\end{eqnarray}
where $v = (p, q, \beta )$ is the set of variables. Contrary to the
earlier case there is only one constraint, given by the second
equation in the above set. It is now straightforward to reduce the
degrees of freedom by mimicing the previous steps.

  An interesting feature is the crucial role played by the first
order equations which, in a conventional analysis \cite{1,2,3}, are always taken
as constraints. Here, on the contrary, if such equations occur in a
second order system (\ref {1}) these are regarded as constraints
whereas, in a first order system (\ref {2}), these are true equations
of motion. The latter interpretation is also valid even in those
cases where a first order system occurs as a subsystem of a second
order system as, for example, happens with the matter sector in
spinor electrodynamics. Incidentally, the inappropriateness of
considering any first order equation as a constraint was also
observed in the symplectic viewpoint \cite{8}.

  To illustrate the above ideas in a simpler setting consider a
nondegenerate singular theory so that the variable $\beta $ is
absent. This corresponds to a second class theory in Dirac's \cite{1}
nomenclature. A typical example is provided by the Proca model,
\begin{equation}
{\cal L} =- \frac{1}{4} F_{\mu\nu}F^{\mu\nu} + \frac{m^2}{2} A_\mu
A^{\mu} \label{3}
\end{equation}
The equations of motion are,
\begin{equation}
\partial_{\mu}F^{\mu\nu} + m^2 A^\nu = 0 \label{4}
\end{equation}
whose zero component is the constraint,
\begin{equation}
(\partial^2 -m^2)A_0 - \partial_0 (\partial_i A_i) = 0 \label{5}
\end{equation}
Time derivative of this constraint vanishes by virtue of the equation
of motion revealing the internal consistency of the model. It is
clear that $A_0$ gets identified with $q$ (\ref {1}). Eliminating 
$A_0$ from (\ref {3}) by using (\ref {5}), the unconstrained
lagrangian is obtained,
\begin{equation}
{\cal L}=- \frac{1}{4} F_{ij}^2 +\frac{1}{2} {\dot A_i}^2 + \frac{1}{2}
\partial_i{\dot A_i}\frac{1}{\partial^2 -m^2}\partial_j{\dot A_j} -\frac{m^2}{2}A_i^2 \nonumber
\end{equation}
The reduced hamiltonian, derived by a standard Legendre transform
from the above lagrangian, is given by,
\begin{equation}
H = \frac{1}{2}\int d^3x (\pi_i^2 +\frac{1}{m^2}(\partial_i\pi_i)^2 +
m^2 A_i^2 +\frac{1}{2} F_{ij}^2) \nonumber
\end{equation}
where $(\pi_i, A^i)$ are the canonical variables. It reproduces the
expression obtained from the Dirac anlysis of eliminating second
class constraints by Dirac brackets and showing that these brackets
reduce to the Poisson brackets for the canonical variables \cite{3}. All these
details are unnecessary in the present context. Moreover explicit
conversion of (\ref {3}) to a first order form, as is required in the
symplectic approach \cite{8}, is avoided.

  Next consider the more interesting case of a degenerate singular
theory, a classic example of which is spinor electrodynamics,
\begin{equation}
{\cal L} =- \frac{1}{4} F_{\mu\nu}F^{\mu\nu} + \bar \psi (i\partial
\!\!\!/\, -m
- eA\!\!\!\!/\,)\psi \label {6}
\end{equation}
The equations of motion are,
\begin{eqnarray}
(i\partial\!\!\!/\,-m -eA\!\!\!\!/\,)\psi &=& 0 \label {7} \\
\partial^\alpha F_{\alpha\mu}- j_\mu &=& 0 \label {8}
\end{eqnarray}
where $ j_\mu =e \bar \psi \gamma_\mu \psi$ is the current. Although
(\ref {7}) is first order it is not a constraint since the matter
sector is first order. Only the $\mu =0$ component of (\ref {8}) is a
constraint. Furthermore there is a degeneracy in these equations
which follows from current conservation. As before, the multiplier
$A_0$ (identified with $q$) can be eliminated in favour of the other
variables by solving the constraint. Using this, (\ref {6}) is
expressed in terms of the reduced set of variables. The Lagrange
equations in these variables are,
\begin{equation}
\partial^j F_{ji} + \partial_0^2[(\delta_{ij} -
\frac{\partial_i\partial_j}{\partial^2})A_j] +
\frac{\partial_i}{\partial^2} \partial_0 j_0 - j_i = 0 \label{9}
\end{equation}
It is obvious that the variable $\beta$, manifesting the degeneracy,
is just the longitudinal $(L)$- component of $A_i$, since it has dropped
out from (\ref {9}). Consequently by choosing the orthogonal polarisation
$A_i = A_i^T + A_i^L$ the lagrangian gets further reduced,
\begin{equation}
{\cal L} = \frac{1}{2} \dot A_i^{T2} - \frac{1}{4} F_{ij}^2 (A^T) +
\frac{1}{2} j_0\frac{1}{\partial^2}j_0 + j_i A_i^T +{\cal L}_M
\label{10} 
\end{equation}
where, expectedly, $A_i^L$ gets automatically removed and ${\cal
L}_M$ is the pure matter part. Denoting the two independent
components of $A_i^T$ by $a_I (I=1, 2)$;
\begin{equation}
A_i^T = (\delta_{iI} - \delta_{i3}\frac{\partial_I}{\partial_3}) a_I \label{11}
\end{equation}
the lagrangian (\ref {6}) is finally expressed in terms of the
independent unconstrained variables $(a_I, \psi )$. The reduced
(physical) hamiltonian, obtained by taking a Legendre transform of
this lagrangian, is given by,
\begin{equation}
H = \int d^3x \{\frac{1}{2}[(\delta_{iI}
 - \frac{\partial_i\partial_I}{\partial^2})p_I]^2 + \frac{1}{4} F_{ij}^2
(a) - \frac{1}{2} j_0\frac{1}{\partial^2}j_0 - j_I a_I +
j_3\partial_3^{-1} \partial_I a_I\} + H_M \label{12}
\end{equation}
directly in terms of the independent canonical pairs $(a^I, p_I),
(\psi, \psi^*)$, and $H_M$ is the pure matter contribution. This
reduced space coincides with that obtained in the Hamiltonian
formalism of abstracting the canonical set by a Levi-Civita
transformation and then evaluating the total hamiltonian on the
constraint surface \cite{3}.

   An analogous treatment, which will also illuminate the connection
with the symplectic approach \cite{8} based on the Darboux transformation, is
now given by converting (\ref {6}) to a first order form,
\begin{equation}
{\cal L} =- \frac{1}{4} F_{ij}^2 -\frac{1}{2} \pi_i^2
- \pi_i F_{0i} + \bar \psi(i\partial\!\!\!/\, -m
- eA\!\!\!\!/\,)\psi \nonumber
\end{equation}
Proceeding as discussed for general first order systems (\ref {2})
one solves the constraint and arrives at a reduced lagrangian,
\begin{equation}
{\cal L} = -\pi_i^T \dot A_i^T -\frac{1}{2} {\pi_i^T}^2 
- \frac{1}{4} F_{ij}^2 (A^T) +
\frac{1}{2} j_0\frac{1}{\partial^2}j_0 + j_i A_i^T +{\cal L}_M
\label{13}
\end{equation}
It is now essential to abstract the {\it independent} canonical pairs
from $(\pi_i^T, A_i^T)$ to obtain the final reduced space. A possible
way is to choose an arbitrary polarisation for these variables from
which the reduced hamiltonian may be derived by a Lagendre transform.
This circuitous path is avoided by making the Darboux
transformation, which comprises (\ref {11}) together with,
\begin{equation}
\pi_i^T =  (\delta_{iI}
 - \frac{\partial_i\partial_I}{\partial^2})p_I \nonumber
\end{equation}
so that the canonical 1-form remains diagonalised $(i.e. \pi_i^T \dot A^{iT}
= p_I \dot a^I)$ and the reduced hamiltonian directly read off from (\ref {13})
reproduces (\ref {12}).

    Finally the issue of gauge fixing is considered with a view to
reveal the subtleties in the simultaneous implementation of $A_3 =0$
and $A_0 =0$ in pure electrodynamics \cite{3,4,9,10}.
Before that it is worthwhile to
point out that the Coulomb gauge $\partial_i A_i =0$ is the most natural
choice since it implies the removal of $A_i^L$ which was gauge independently
identified as the redundant $(\beta)$ variable. Effectively, therefore,
choosing the Coulomb gauge is like not choosing any gauge. Now confining
our attention to the pure Maxwell theory, it is found from (\ref {8}) that
the solution for the multiplier consistent with the axial $(A_3 =0)$ gauge
is,
\begin{equation}
A_0 = \frac{\partial_0}{\partial^2} (\partial_I A_I); I=1, 2 \label{B}
\end{equation}
Eliminating both $A_3$ and $A_0$ from (\ref {6}) yields the unconstrained
lagrangian which, after the usual Legendre transform, leads to the reduced
hamiltonian,
\begin{equation}
H_{axial} = \int d^3x \frac{1}{2}[ (\frac{\partial_I}{\partial_3} \pi_I)^2
+ \pi_I^2
- A_I \partial^2 A_I - (\partial_I A_I)^2 ] \label{A}
\end{equation}
with $(A_I, \pi^I)$ being the canonical pairs. It is simple to check that
the canonical transformation,
\begin{equation}
\pi_I = -\sqrt {-\partial^2}a_I; A_I = \frac{1}{\sqrt {-\partial^2}} p_I
\nonumber
\end{equation}
establishes the equivalence of (\ref {A}) with the gauge independent result
(\ref {12}). Hence the axial gauge supplemented with (\ref {B}) is an
"allowed" choice. If, on the contrary, $A_0 =0$ is taken instead of
(\ref {B}) then the reduced hamiltonian is now given by (\ref {A}), but
without its first term. Hence canonical equivalence with (\ref {12})
cannot be shown so that the choice $A_0 =0$ with $A_3= 0$ is disallowed.

The picture is different in the hamiltonian analysis where the familiar
expression for the total hamiltonian is given by,
\begin{equation}
H_T = \int d^3x (\frac{1}{2}\pi_i^2 +\frac{1}{4}F_{ij}^2 + A_0\partial_i\pi_i
+\lambda\pi_0) \label{C}
\end{equation}
If the constraints $\pi_0 =0, \partial_i\pi_i =0$ are implemented by fixing
$A_0=0$ and $A_3=0$, then the reduced hamiltonian obtained from (\ref {C})
exactly corresponds to (\ref {A}). Moreover the Dirac brackets of $A_I, \pi^I$
are identical to their Poisson brackets so that these variables constitute the
canonical set \cite{2,3}. Contrary to the lagrangian formulation, therefore, the
axial-temporal gauge is a valid choice in the hamiltonian framework. Moreover
the axial gauge imposed together with $\partial_3 A_0 -\pi_3 =0$, which is the
hamiltonian analogue of (\ref {B}), also yields a valid reduced space \cite{2,3}. 
Consequently the hamiltonian formalism admits a wider class of admissible
gauges than the lagrangian formalism.

   It is clear that the practical viability of this scheme depends on
solving the constraint. While this can always be done in abelian theories
(including Chern-Simons terms) and the redundant $(\beta)$ variable
identified with $A_i^L$, the same cannot be said for nonabelian theories
involving nonlinear constraints. This difficulty, it is emphasised,
is more technical than conceptual. The present method, however, suggests
an intriguing possibility of solving the nonlinear constraint and identifying
$\beta$ by a perturbative series around the known abelian results. It should
be mentioned that even in the hamiltonian formulation it is problematic
to generalise the Levi-Civita \cite{5}
method to systematically reduce the degrees of
freedom in a nonabelian theory without gauge fixing \cite{4,6}.
Nevertheless, in those
cases where a gauge independent reduction is constructively feasible, this
approach provides a definite simplification over the elaborate hamiltonian
formalism based on the Levi-Civita transformation \cite{4,5,6,7}.
Furthermore, by discussing
both second and first order systems within a unified framework, it dispenses
with the Darboux transformation \cite{8}. 

    This work was supported by CNPq-Brazilian National Research Council. The author also thanks members of the IF, UFRJ for their kind hospitality.


\begin{thebibliography}{99}
\bibitem{1} P.A.M. Dirac, "Lectures on Quantum Mechanics" (Belfer Graduate
School of Science, Yeshiva Univ. New York, 1964.)

\bibitem{2} A. Hanson, T. Regge and C. Teitelboim, "Constrained Hamiltonian
Systems" (Academia Nazionale dei Lincei, Rome, 1976.)

\bibitem{3} D. Gitman and I. Tyutin, "Quantisation of Fields with
Constraints" (Springer- Verlag, 1990)

\bibitem{4} S. Gogilidze, A. Khvedelidze and V.Pervushin, Phys. Rev. D53, 2160 (1996).

\bibitem{5} T.Levi-Civita, Prace Mat.-Fiz 17, 1 (1906).

\bibitem{6} S.Shanmugadhasan, J. Math. Phys. 14, 677 (1973).

\bibitem{7} J.Goldberg, E.Newman and C.Rovelli, J. Math. Phys. 32, 2739 (1991);
Also see D.Boyanovsky, E.Newman and C.Rovelli, Phys. Rev. D45, 1210 (1992).

\bibitem{8} For a modern perspective, see R.Jackiw in "Diverse Topics in
Theoretical and Mathematical Physics" (World Scientific, Singapore, 1995).

\bibitem{9} R.Sugano and T.Kimura, J. Phys. A16, 4417 (1983) and references
therein.

\bibitem{10} M.Lavelle and D.McMullan, Phys. Lett. B316, 172 (1993).

\end{thebibliography}
\end{document}